\documentclass[%
superscriptaddress,
reprint,
amsmath,amssymb,
aps,
prx,
10pt,
twocolumn,
floats,
]{revtex4-2}

\usepackage[english]{babel}
\usepackage[utf8]{inputenc}
\usepackage[T1]{fontenc}
\usepackage{graphicx}
\usepackage[caption=false]{subfig}
\usepackage{float}
\usepackage{dcolumn}
\usepackage{blindtext}
\usepackage{upgreek}
\usepackage{siunitx}
\usepackage{xcolor}

\DeclareSIUnit\gauss{G}
\usepackage{braket}

\usepackage{wasysym}
\usepackage{enumitem}
\usepackage{bm}

\usepackage{amsmath}
\usepackage{multirow} % For the \multirow command
\usepackage{siunitx}  % For the \SI and unit formatting
\usepackage{array}    % For advanced table column control

\begin{document}

\title{Full-field-of-view aberration correction for large arrays of focused beams}

\author{Y.~Machu}
\author{G.~Creutzer}
\affiliation{Laboratoire Kastler Brossel, Coll\`ege de France, CNRS, ENS-Universit\'e PSL, Sorbonne Universit\'e, 11 place Marcelin Berthelot, F-75231 Paris, France}
\author{C.~Sayrin}
\affiliation{Laboratoire Kastler Brossel, Coll\`ege de France, CNRS, ENS-Universit\'e PSL, Sorbonne Universit\'e, 11 place Marcelin Berthelot, F-75231 Paris, France}
\affiliation{Institut Universitaire de France, 1 rue Descartes, 75231 Paris Cedex 05, France}
\author{M.~Brune} 
\email[Corresponding author: ]{michel.brune@lkb.ens.fr}
\affiliation{Laboratoire Kastler Brossel, Coll\`ege de France, CNRS, ENS-Universit\'e PSL, Sorbonne Universit\'e, 11 place Marcelin Berthelot, F-75231 Paris, France}

\date{\today}

% use {asbstract*} to suppress the copyright line. Copyright information will be added in production

\begin{abstract} 
We implement an aberration correction method for the creation of extended arrays of spots well beyond the isoplanatic region of any optical system. The method relies on an extensive calibration of aberrations in terms of Zernike polynomials over the full accessible field of an optical system. We use a modified Gerchberg-Saxton algorithm for generating holographic phase masks creating fully corrected arbitrary arrays of spots. By applying the method to an aspherical lens, and using a liquid-crystal spatial light modulator (SLM), we increase the aberration-free field of view from 50 to 500 $\mu$m, limited by the largest diffraction angles accessible to the SLM. This opens new perspectives for the generation of large arrays of optical tweezers, when the field of view becomes limited by aberrations, especially for neutral atom-based quantum processors and simulators.
\end{abstract}

\maketitle

%%%%%%%%%%%%%%%%%%%%%%%%%%  body  %%%%%%%%%%%%%%%%%%%%%%%%%%
\section{Introduction}

Control of optical aberrations plays an essential role in designing optical systems to reach diffraction-limited performances\cite{born_principles_1993}. It is of paramount importance for large-field-of-view imaging with high-numerical-aperture lenses\cite{booth_adaptive_2014, zheng_characterization_2013} and to create arrays of optical tweezers, with applications in biology\cite{nilsson_review_2009, schermelleh_guide_2010, dholakia_shaping_2011, wu_speed_2021, yang_optical_2021}, parallel laser micro-machining\cite{hasegawa_adaptive_2009, hasegawa_massively_2016, zhang_-system_2020, wang_femtosecond_2021, kuroo_holographic_2025, mauclair_dynamic_2025} and micromanipulation\cite{grier_revolution_2003, bianchi_real-time_2010, yang_optical_2021}, parallel 3D printing\cite{jesacher_parallel_2010, zhang_high-throughput_2024, somers_physics_2024} and trapping of single atoms\cite{nogrette_single-atom_2014, barredo_synthetic_2018, kim_large-scale_2019, schymik_situ_2022}, with recent developments that led to the creation of large arrays of tweezers trapping thousands of atoms\cite{manetsch_tweezer_2025, pichard_rearrangement_2024, chiu_continuous_2025, lin_ai-enabled_2025}. A single lens can be used to reach the diffraction limit on its optical axis even with a large numerical aperture by using an aspherical surface that suppresses the spherical aberration. However, this simple system has a limited field of view (FOV) due to growing aberrations away from the optical axis. A standard solution consists in combining many lenses whose individual aberrations cancel each other in optimized combinations. Nevertheless, these assemblies, which often contain ten to twenty lenses, are complex, bulky and expensive.

Beyond design optimization, adaptive optics\cite{babcock_possibility_1953,booth_adaptive_2002, davies_adaptive_2012, booth_adaptive_2014, salter_adaptive_2019} has become a popular method of aberration compensation. It relies on an estimate of the wavefront distortion and on a feed-back correction by a spatial light modulator, such as a deformable mirror or an array of liquid crystals (SLM). This technique enables perfect compensation of geometrical aberrations and the creation of a diffraction-limited spot centered on any given point within the accessible field of view (FOV) of an optical system. However, the spatial dependency of the aberrations limits the efficiency of this method to a restricted isoplanatic area of the order of the native diffraction-limited FOV of the optical system\cite{fried_anisoplanatism_1982, simmonds_modelling_2013, mertz_field_2015}. Therefore, local corrections considerably increase the usable FOV but they do not enable the production of an aberration-free image over the whole accessible field. In the case of aberrations originating from propagation over large distances before entering the optical system, e.g. in the atmosphere for astrophysics, this limitation can be overcome using a multiconjugate adaptive optics method significantly increasing the diffraction limited field of view\cite{rigaut_multiconjugate_2018}. For the generation of corrected arrays of spots, a zonal approach can be used to correct individual spots with a spatial light modulator (SLM), at the expense of a reduced effective numerical aperture\cite{park_large-field--view_2017-2,pathak_zonal_2025-1}. In the context of parallelized scanning microscopy for high resolution imaging of biological samples, a general method for correcting simultaneously a large number of spots without affecting the numerical aperture has been introduced in Ref.\cite{pozzi_anisoplanatic_2020}. The method relies on the holographic correction of aberrations using a SLM. 

In this paper, we propose and apply a method similar to that introduced in\cite{pozzi_anisoplanatic_2020} to create  arbitrary arrays of diffraction-limited spots in the focal plane of a centered optical system. It relies on SLM-controlled holography to simultaneously correct the aberrations for all the spots in the array, covering the full accessible FOV. We model the variation of aberrations within the full accessible field by introducing modified Seidel coefficients as a variant of the well-known Seidel expansion describing the field dependency of the wavefront distortion introduced by the optical system\cite{machu_patent_nodate, machu_mesure_2024}. Our aberration compensation method relies on a preliminary measurement of only a few modified Seidel coefficients, which can then be used to generate arbitrary arrays of spots with an arbitrary shape. As a demonstration, we apply the method to the correction of a low-cost, aspherical lens with a native FOV limited to $50\times 50\,\mu$m. After correction, we generate large, diffraction-limited arrays of up to 5194 spots in a field of $500\times 500\,\mu$m, limited by the maximum diffraction angle of our SLM. We fully characterize the optical quality of the spots and we demonstrate a tenfold increase of the size of the diffraction-limited field of view.

\section{Principle of the method}

 \begin{figure}[htbp]
\centering\includegraphics[width=8cm]{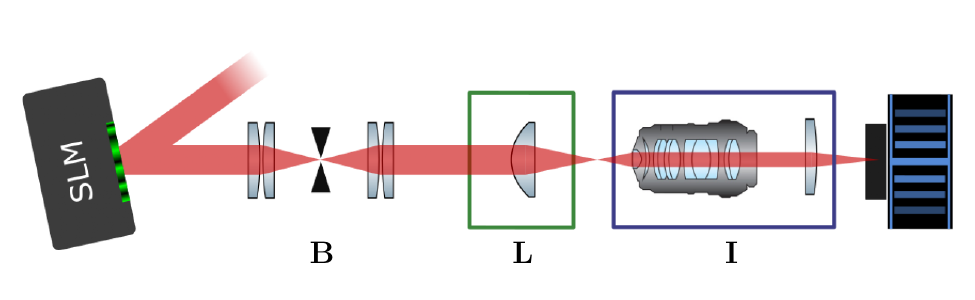}
\caption{ Experimental setup.  A collimated input laser beam is diffracted using the SLM. We form the image of the SLM on the back focal plane of the convergent optical system of interest $\bf L$ (focal distance $f$) by using an afocal telescope in the 4f configuration. A blade {$\bf B$} removes the zero-order reflection. An array of spots is formed in the focal plane of $\bf L$. We test the quality of this array by imaging with a camera using an aberration-free optical system $\bf I$ described in the supplemental document.}
\label{Fig1}
\end{figure}

The principle of our SLM-based optical setup is depicted on fig.\,\ref{Fig1}. A lens $\bf L$ of focal length $f$ creates in its focal plane an array of focused beams with a possibly complex intensity pattern (Bottle beams\cite{arlt_generation_2000,zhang_magic-wavelength_2011} or Laguerre-Gaussian beams for instance). The SLM (Hamamatsu X13138-03) is a liquid crystal (LCoS) matrix used as a phase-only modulator. It applies an arbitrary phase mask $\phi({\bm \rho})$ on the incident wave (wave number $k$). A 4f telescope generates in the back focal plane of $\bf L$ an optical field of amplitude $E({\bm \rho})= \sqrt{I({\bm \rho})} e^{i\phi({\bm \rho})}$ where ${\bm \rho}$ is the position coordinate in this plane, with the origin set on the optical axis of $\bf L$ and $I({\bm \rho})$ is the intensity profile of the beam. In the focal plane of an ideal lens, the field amplitude $F(\bm r)$ is the Fourier transform of $E({\bm \rho})$. The origin of coordinate $\bm r$ is set on the optical axis of $\bf L$, too.

For the generation of an array of $N$ spots at positions $\{\bm{r}_p\}_{p\in \{1,..., P\}}$ in the focal plane of $\bf L$, the theoretical field amplitude in the back focal plane of an ideal lens reads
\begin{equation}
E_{\mathrm{th}}({\bm \rho})= \sqrt{I({\bm \rho})}\, a_{\mathrm{th}}({\bm \rho}),
\label{Exact}
\end{equation}
where $a_{\mathrm{th}}({\bm \rho})=\sum_p w_p.e^{i {k\over{f}}\bf{r}_p\cdot{\bm \rho}}$ is a superposition of plane waves and $\{w_p\}$ are complex weights. This amplitude generates an array of spots with an intensity proportional to $|w_p|^2$. The shape of individual spots is determined by $I(\rho)$, which is supposed to have a slow variation at the scale of the pupil defined by the SLM. As the dimensionless amplitude $a_{\mathrm{th}}({\bm \rho})$ does not have a uniform modulus, it cannot be generated with a phase-only modulator. One usually uses an iterative Weighted Gerchberg-Saxton (WGS) algorithm\cite{gerchberg_practical_1972, leonardo_computer_2007, barredo_synthetic_2018, kim_large-scale_2019} to compute a phase mask, that approximately generates the target array of spots. The algorithm uses the following ansatz for the phase $\phi_{\mathrm{nc}}({\bm \rho})$ of the non-corrected field amplitude in the back focal plane of $\bf L$
\begin{equation}
\phi_{\mathrm{nc}}({\bm \rho})=\text {arg}\bigg(\sum_p w_p\cdot e^{i {k\over{f}}\bf{r}_p\cdot{\bm \rho}}\bigg),
\end{equation}
which is the phase at position ${\bm \rho}$ of the ideal amplitude $a_{\mathrm{th}}({\bm \rho})$. The algorithm iterates between the back focal plane and the Fourier plane of an ideal lens by direct and inverse Fourier transforms. It finds complex amplitudes $w_p$'s that make the diffraction pattern of $E_{\mathrm{nc}}(\bm \rho)=\sqrt{I({\bm \rho})}\, e^{i\phi_{\mathrm{nc}}({\bm \rho})}$ close to that of $E_{\mathrm{th}}(\bm \rho)$. 

In practice, the lens $\bf L$ introduces aberrations, which are not taken into account by the simple Fourier transform used in a standard WGS algorithm. The approximate non-corrected phase mask $\phi_{\mathrm{nc}}({\bm \rho})$ thus produces arrays of deformed spots whose shape depends on their position in the FOV.

{\bf Field-dependent aberration correction method}. 
Let us first consider a single-spot target image at position $\bm r_p$ corresponding to the input amplitude $E_p(\bm \rho)=\sqrt{I({\bm \rho})}e^{i {k\over{f}}\bm{r}_p\cdot{\bm \rho}}$. The aberrations introduced by the non-ideal system $\bf L$ can be represented by a phase distortion mask  $\Phi({\bm r}_p, {\bm \rho})$ producing the amplitude $E'_p(\bm \rho)=E_p(\bm \rho)e^{i\Phi({\bm r}_p, {\bm \rho})}$ when applied to the incident plane wave. When focused by an ideal, i.e., aberration-free lens, this amplitude produces a spot at position ${\bm r}_p$ with the same aberrations as those of the actual lens. The corresponding amplitude in the focal plane is the Fourier transform of $E'_p(\bm \rho)$ instead of that of $E_p(\bm \rho)$. For an arbitrary input amplitude $E(\bm \rho)$ the  amplitude $F(\bm r)$ in the focal plane of $\bf L$, including the effect of aberrations, is
\begin{equation}
F(\bm r)= \iint e^{-i {k\over{f}}\bm{r}\cdot{\bm \rho}}\Big\{E(\bm \rho)\,e^{i \Phi({\bm r}, {\bm \rho})}\Big\} d^2\rho.
\label{FTphi}
\end{equation}
The function $\Phi({\bm r}, {\bm \rho})$ represents a complete description of geometrical aberrations in the full optical field of $\bf L$. Note that due to the $\bm r$ dependency of $\Phi({\bm r}, {\bm \rho})$ equ.\,\eqref{FTphi} is not a Fourier transform.

The adaptive optics correction approach\cite{booth_adaptive_2014, salter_adaptive_2019} consists in applying with the SLM the phase mask $-\Phi({\bm r}_p, {\bm \rho})$ to the amplitude $E_p(\bm \rho)$ in order to compensate exactly the aberrations of an image centered at ${\bm r}_p$. The variation of  $\Phi({\bm r}, {\bm \rho})$ as a function of ${\bm r}$ defines the diffraction-limited field of view of $\bf L$. It is usually restricted to the vicinity of the corrected position ${\bm r_p}$. For this position, the phase-distortion mask $\Phi({\bm r}_p, {\bm \rho})$ can be measured as a function of $\bm \rho$ by using a phase reconstruction method based on the Gerchberg-Saxton algorithm\cite{hanser_phase_2003, barredo_synthetic_2018} (see supplemental document). In the following, we assume that a mask $\phi_0(\bm \rho)$ exactly compensating aberrations at $\bm r=0$ has been measured and applied with the SLM in order to cancel aberrations locally, close to the optical axis of the lens. Note that the application of the mask $\phi_0(\bm \rho)$ is equivalent to the standard adaptive optics approach used for correcting a single point. The residual field-dependent aberration of this on-axis pre-compensated system are thus described by $\Phi(\bm r, {\bm \rho})$ with $\Phi(0, {\bm \rho})=0$. 

We now present our method for the full-field correction of the off-axis aberrations.
It first consists in slightly modifying the WGS ansatz using
\begin{equation}
\phi_c({\bm \rho})=\mathrm{arg}\bigg(\sum_p w_p\,e^{i {k\over{f}}\bf{r}_p\cdot{\bm \rho}}\,\, e^{-i \Phi({\bm r}_p, {\bm \rho})}\bigg).
\end{equation}
in place of $\phi_\mathrm{nc}({\bm \rho})$. The corrected ansatz amounts to applying to each ${\bm k}_p$ component of the sum a phase mask undoing the aberrations that will be introduced by $\bf L$. We then introduce a corrected WGS algorithm, which additionally account for aberrations in the calculation of the field amplitude in the Fourier plane of $\bf L$ according to equ.\,\eqref{FTphi}. In this way, we can generate phase masks correcting all the spots of an array in a large FOV. The CWGS algorithm has similar convergence as the standard WGS one. To the best of our knowledge, this CWGS was first introduced in\cite{machu_patent_nodate, machu_mesure_2024}.

The application of our correction method relies on a precise determination of $\Phi({\bm r},{\bm \rho})$. A first approach consists in determining this function sequentially by successively applying with the SLM the phase gradient generating one spot at ${\bm r}_p$. For this single spot, $\Phi({\bm r}_p, {\bm \rho})$ can be determined with the same phase reconstruction method  \cite{hanser_phase_2003} as used for the determination of $\phi_0(\bm \rho)$. However, the determination of $\Phi({\bm r},{\bm \rho})$ at every spot of the final image is time-consuming, and becomes impractical for the generation of large arrays. Alternatively, we will now show that one can use a simple model of the variations of $\Phi({\bm r}, {\bm \rho})$ as a function of ${\bm r}$ involving only a few parameters, which, once measured, can be used to generate arbitrary corrected arrays without the need for additional measurements.

{\bf Simple model of $\Phi({\bm r}, {\bm \rho})$}. Following the approach introduced by Seidel to describe the field-dependent aberrations of a centered optical system, we start by a Taylor expansion of $\Phi({\bm r}, {\bm \rho})$ as a function of ${\bm r}$ and ${\bm \rho}$\cite{born_principles_1993}. Due to axial symmetry, all terms with odd global polynomial order in ${\bm r}$ and ${\bm \rho}$ vanish. We rewrite the Taylor expansion in terms of Zernike polynomials $Z_{n}^{m}({\bm \rho})$, with  $n \in \mathbb{N}$, $m \in \mathbb{Z}$ and $|m| \le n$\cite{conforti_zernike_1983, lakshminarayanan_zernike_2011}. To the lowest-order terms relevant to describe aberrations, this expansion can be written in terms of the modified Seidel coefficients $C'$,  $C''$, $A'$, $K'$ and $D'$:
\begin{equation}
	 \begin{split}
		\Phi(\bm{r},\bm{{\bm \rho}}) &=  C^{\prime}\left[x \,Z_{3}^{1}(\bm{{\bm \rho}})+ y \,Z_{3}^{-1}(\bm{{\bm \rho}})\right] \\
		&+ C'' \left[ x\,Z_5^1(\bm{\rho}) + y\, Z_5^{-1}(\bm{\rho}) \right]\\
		&+ A^{\prime} \left[(x^{2}-y^{2}) \,Z_{2}^2(\bm{{\bm \rho}})+ 2xy \,Z_{2}^{-2}(\bm{{\bm \rho}})\right] \\
		&+ K^{\prime}(x^{2}+y^{2}) \,Z_{2}^0(\bm{{\bm \rho}}) \\ 
		&+D^{\prime}\left[(x^{2}+y^{2})x\,Z_{1}^1(\bm{{\bm \rho}})+(x^{2}+y^{2})y\,Z_{1}^{-1}(\bm{{\bm \rho}})\right]+ ...\ ,
	\end{split}
\end{equation}
where $x$ and $y$ are the cartesian components of ${\bm r}$. The modified Seidel coefficients $C'$, $C''$, $A'$, $K'$ and $D'$ describe the primary and secondary coma, the astigmatism, field curvature and distortion, respectively. As a result of pre-compensation using the phase mask $\phi_0(\bm \rho)$, all these aberrations vanish on the optical axis and have a polynomial dependency on $\bm r$. The modified Seidel coefficients are linear combinations of the standard ones. The former can be more easily measured by the method described below, which benefits from the orthogonality of Zernike polynomials. 

\section{Experimental characterization of field-dependent aberrations}

We apply our aberration compensation method to a standard aspherical lens (Asphericon AFL12-15-P-U-285). It is optimized at the wavelength of 285\,nm. Its measured Strehl ratio is $0.7$ on the optical axis at our working wavelength of 821\,nm, with a numerical aperture of 0.37 and a focal length $f=16.32\,$mm. The imaging system ($\bf I$ in fig.\,\ref{Fig1}) is a combination of microscope objectives described in the supplemental document. We have carefully checked that it introduces negligible coma, astigmatism and curvature into the full field. The SLM is illuminated by a Gaussian beam of waist 5.2\,mm, and we apply a pupil of 12\,mm diameter on the SLM.

To characterize field aberrations, we successively measure the wavefronts $\phi_q({\bm \rho})=\Phi({\bm r_q}, {\bm \rho})$ for single spots located at positions ${\bm r}_q$ of a $7\times7$ square reference array centered on the optical axis.
The reference array covers a square field of $400\times 400\,\mu$m. By defining the $\theta_n^m({\bm r}_q)$ coefficients as the scalar product between $\phi_q({\bm \rho})$ and  the Zernike polynomial $Z_n^m({\bm \rho})$ we get
\begin{equation}
\phi_q({\bm \rho})= \sum_{n,m} \theta_n^m({\bm r}_q)\,Z_n^m({\bm \rho}).
\label{phi}
\end{equation}
We obtain the coefficients $C'$, $C''$, $A'$ and $K'$ with a fit to the measured values of the corresponding $\theta_n^m({\bm r}_q)$, with a linear function of $\bm r$ for the coma and quadratic functions of $\bm r$ for the astigmatism and curvature. For the quadratic fits of astigmatism and curvature, we also take as free parameters the centers of the paraboloids and take them into account in the correction process. It amounts to introducing odd-order terms in the Taylor expansion of $\Phi({\bm r}, {\bm \rho})$. This is a heuristic approach, which accounts for imperfect axial symmetry of the setup and improves the quality of the correction. The fit functions are given in the supplemental document. We also measured the distortion coefficient $D'$ but we estimate that it is dominated by that of the imaging system. We thus did not consider the distortion term in the correction process.

\begin{figure*}[htbp]
\centering\includegraphics[width=13cm]{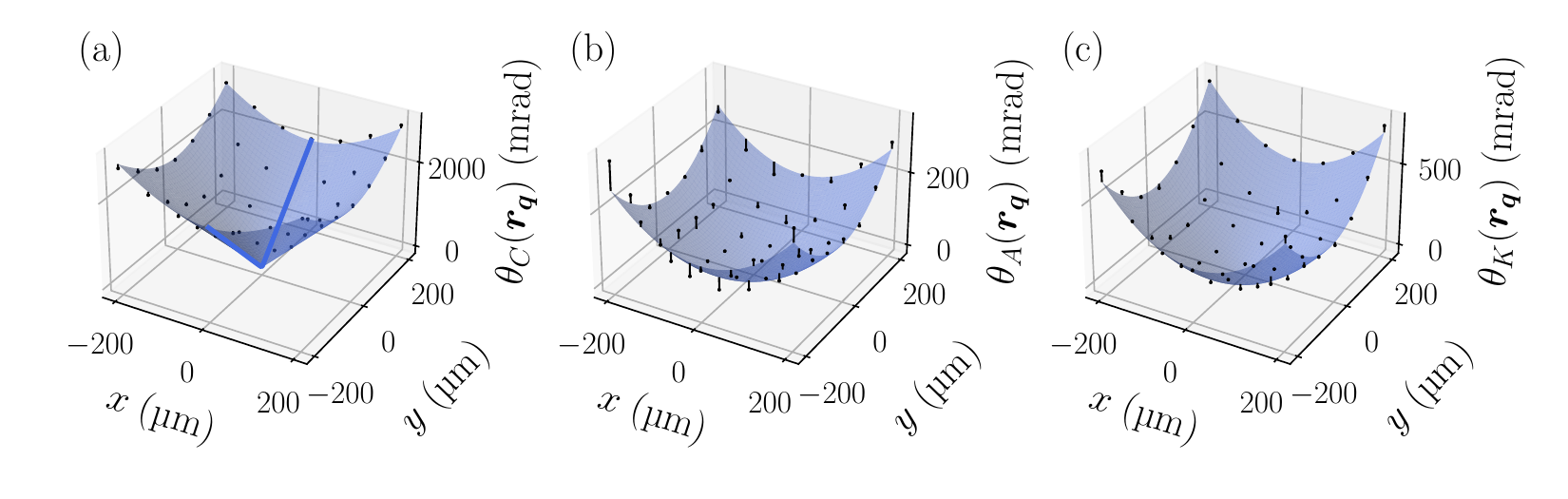}
\caption{ Spatial dependence of the aberrations. (a) First-order coma. The blue line highlights the linear dependence of $\theta_C({\bm r}))$ as a function of $|\bm r|$.  Panel (b) corresponds to astigmatism and (c) to curvature. 
The points are measurement results, while the shaded surfaces correspond to the fits. The drop lines are the residuals of the fits. The paraboloids corresponding to astigmatism and curvature are centered at positions $(-2\mu\text m,-3\mu\text m)$ and $(7\mu\text m,-18\mu\text m)$, respectively. These values are small compared to the dimension of the corrected field. The larger values obtained for the field curvature can be explained by a small tilt between the focal plane of the aspherical lens and the plane of the camera sensor.}
\label{Fig2}
\end{figure*}
We show in fig.\,\ref{Fig2} the variation of the functions $\theta_X({\bm r})=\sqrt{ \theta_n^m({\bm r})^2+ \theta_n^{-m}({\bm r})^2}$ with $(X,n,m)=(C,3,1)$, $(A,2,2)$ and $(K,1,1)$ for coma, astigmatism and curvature, respectively. The linear dependence of the two terms $\theta_3^{\pm 1}(\bm r)$ contributing to coma translates into a linear variation of $\theta_C({\bm r})$  as a function of the distance $|\bm r|$ to the optical axis (fig.\,\ref{Fig2}(a)). The astigmatism and curvature terms (fig. \ref{Fig2}(b-c)) have the expected quadratic dependency with respect to $|\bm r|$. From the linear and quadratic fits, we obtain the values of the modified Seidel coefficients $C'= 10.04 (4)\,$mrad$/\mu$m, $A'= 3.30 (7)\,\mu$rad$/\mu$m$^2$ and $K'= 8.30 (7)\,\mu$rad$/\mu$m$^2$. These values are close to those calculated from the lens geometry using the OSLO software\cite{noauthor_oslo_nodate} presented in table\,1 of the supplemental document. From the measurements, we also obtain the second-order coma coefficient $C''= 0.13 (1)\,$mrad$/\mu$m and take it into account in the aberration correction. Note that $C''$ is two orders of magnitude smaller than $C'$, it has a marginal impact on the final quality of the spots. It was not necessary to include more terms to reach the diffraction limit over a large field of view.

\section{Results}

Using the WGS and CWGS algorithms, we generate two phase masks corresponding to a 34$\times$34 square array with a $d=15\,\mu$m step. Note that the computational overhead of running the CWGS instead of the WGS is not significant. These arrays contain 1156 spots and cover a 495$\times$495$\,\mu$m field of view, close to the maximal value of 535\,$\mu$m corresponding to the Nyquist limit set by the number of pixels of the SLM, the wavelength and the focal length of $\bf L$. The focal plane of the aspherical lens is imaged on the camera using a combination of microscope objectives imaging the full array (see supplemental document).

\begin{figure*}[t]
\centering\includegraphics[width=14cm]{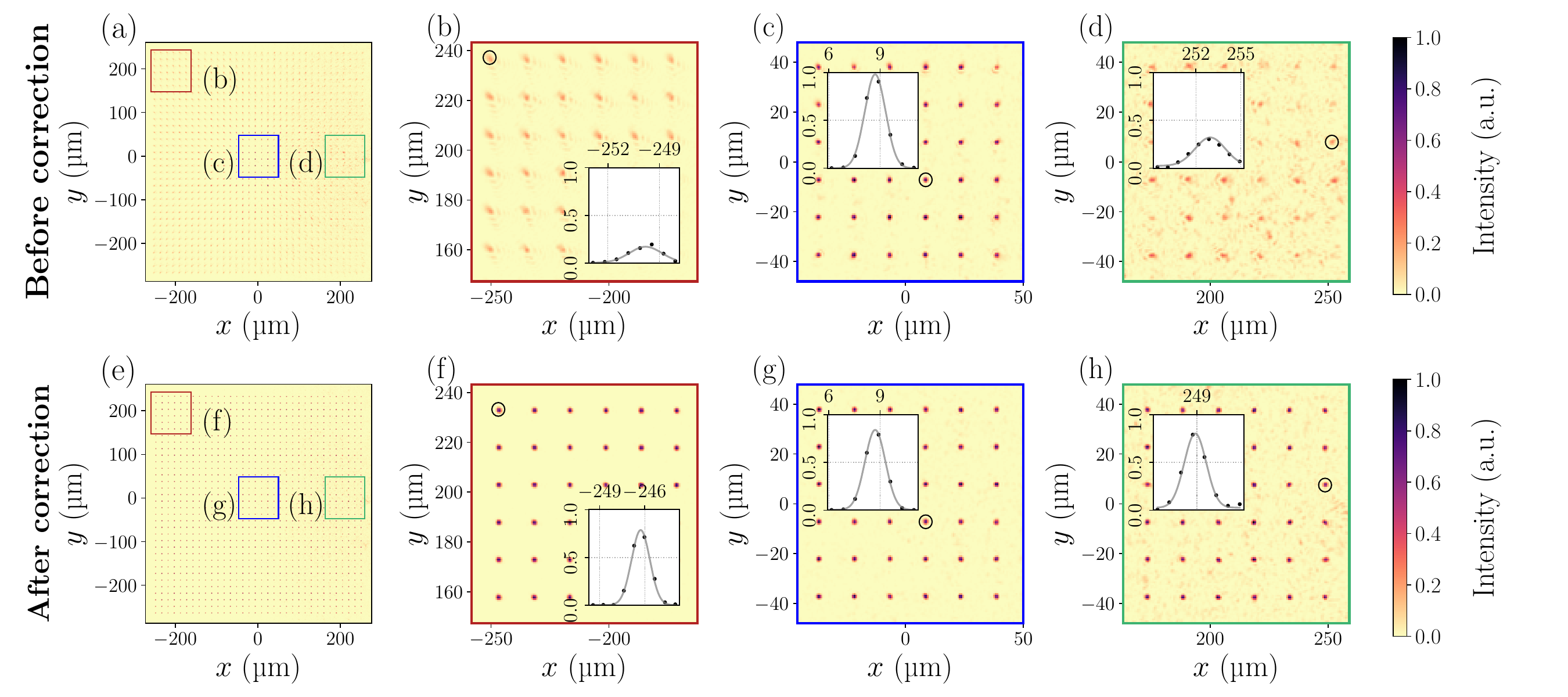}
\caption{ Images of two 34$\times$34 arrays without [WGS, panels (a-d)] and with field aberration correction [CWGS, panels (e-h)]. The six right panels are zooms onto regions of interest defined by colored squares on the corresponding left panel. Intensities are normalized to the maximum intensity of the central spot highlighted by a circle on panel (c) and (g). The insets are cross-sections along the $x$ direction of the intensity profile of the spots highlighted by a black
 circle in each panel. Cross-sections are centered to contain the highest-intensity pixel of the selected spot.}
\label{Fig3}
\end{figure*}

Figure\,\ref{Fig3}(a) shows the full-field image of a non-corrected array generated using the WGS algorithm and including aberration compensation on the optical axis only with the phase mask $\phi_0(\bm \rho)$. Figures\,\ref{Fig3}(b-d) are zooms onto various regions depicted by squares on fig.\,\ref{Fig3}(a). The insets are horizontal cross-sections of the spots highlighted by a circle on the corresponding panel. Panel (c) shows diffraction-limited performance for spots located up to 25 $\mu$m from the optical axis. This defines the native field of view of the aspherical lens. Panel (b) shows a strong distortion of spots, mainly due to coma. Panel (d) is close to the zeroth diffraction order of the SLM located at the right border of the array. In addition to aberrations, it shows significant fluctuations of maximal intensities and an important background speckle. Figures\,\ref{Fig3}(e-h) show similar images obtained after full-field aberration compensation based on the CWGS algorithm. The high quality of spots even close to the border of the field is conspicuous. Spots located at the border of the array have similar maximum intensities and shapes as the central spot depicted in the inset of fig.\,\ref{Fig3}(g).

\begin{figure*}[tp]
\centering\includegraphics[width=13cm]{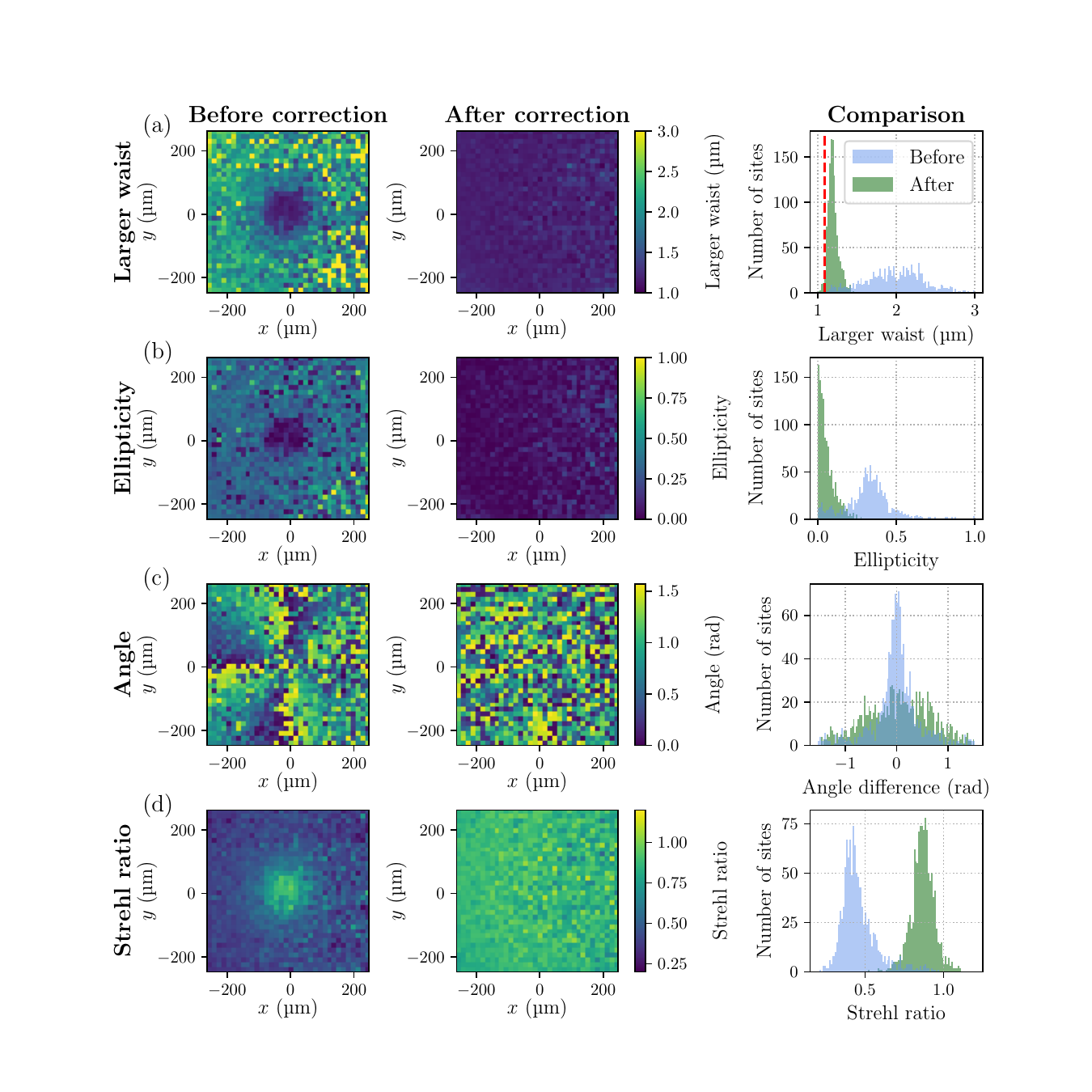}
\caption{Performance of the aberration correction procedure. (a) Spatial variation of the larger waist of spots, determined by fits before (left) and after (center) applying field correction. In both cases, the global correction mask $\phi_0$ corrects aberrations on the optical axis. Right panel: histogram of larger-waist distribution before (blue) and after correction (green). The vertical dotted line corresponds to the expected diffraction-limited waist. (b-d) Similar representations of the ellipticity (b), of the orientation of the major axis of spots (c) and of the Strehl ratio of the spots (d). Note that due to the experimental uncertainty in the determination of the maximum peak intensity, the measured Strehl ratio may be larger than 1.  For the two left  panels of (c) we represent the orientation of the major axis of ellipses relative to the $x$ axis. The right panel of (c) represents the histogram of the orientation of the major axis of ellipses relative to a radial axis.}
\label{Fig4}
\end{figure*}

In order to quantitatively assess the performances of the aberration compensation method, we perform individual two-dimensional Gaussian fits of the focal points including a uniform background intensity (see supplemental document). We determine for each spot the larger waist $w_{\mathrm{max}}$, the ellipticity $1-w_{\mathrm{min}}/w_{\mathrm{max}}$ (with $w_{\mathrm{min}}$ the smaller waist), the orientation of the major axis of the elliptical shape, and the Strehl ratio. The variation of these parameters within the field of view is represented on fig.\,\ref{Fig4}. At the center of the image, the largest waist before correction [fig.\,\ref{Fig4}(a), left panel] is about $1.2\,\mu$m on the optical axis, close to the expected diffraction limit estimated at $1.09(1)\,\mu$m  (see supplemental document). It increases rapidly to about $2\,\mu$m  with a broad distribution (green histogram on right panel) when the distance to the optical axis is larger than $25\,\mu$m.
After field aberration correction, one observes a strong reduction of the largest waist whose distribution (green histogram on right panel) peaks at $1.19(7)\,\mu$m. The standard deviation of the larger-waist distribution is strongly reduced, too, by the aberration correction, from $0.5\,\mu$m to $0.1\,\mu$m. 

Figure\,\ref{Fig4}(b) presents the ellipticity of the spots. Before field correction, most of the spots have an elliptical shape. After field correction, the ellipticity of the spots drops to the negligible average value of 0.07(7), and all spots are nearly circular. Left panel of figure\,\ref{Fig4}(c) shows a clear radial orientation of the major axis of the  elliptical spots observed without field correction. After field correction, the orientation of ellipses is dominated by noise [central panel of figure\,\ref{Fig4}(c)], without any structure even at the border of the field. Figure\,\ref{Fig4}(d) presents the spatial variation of the Strehl ratio, $S$. Without field correction, it defines a diffraction-limited field of view ($S>0.8$) of about $50\,\mu$m diameter with an average value over the full-field $\bar S=0.46$. After field correction, the average value of the Strehl ratio increases to $\bar S=0.87(7)$. One observes a homogeneous performance over the full field of view with a maximal diagonal extension of 700\,$\mu$m, demonstrating the ability of our method to correct the aberrations simultaneously for all the spots within the full accessible optical field. From the fit of individual spots, we observe a reduction of the relative standard deviation of the total power of individual spots from 19\% to 2\% when using the CWGS instead of the WGS algorithm. At the same time, we observe that the average background intensity drops by nearly one order of magnitude when using the CWGS algorithm. Our method thus produces large arrays of spots with high contrast and excellent spot-power homogeneity. 

\begin{figure*}[t]
\centering\includegraphics[width=13cm]{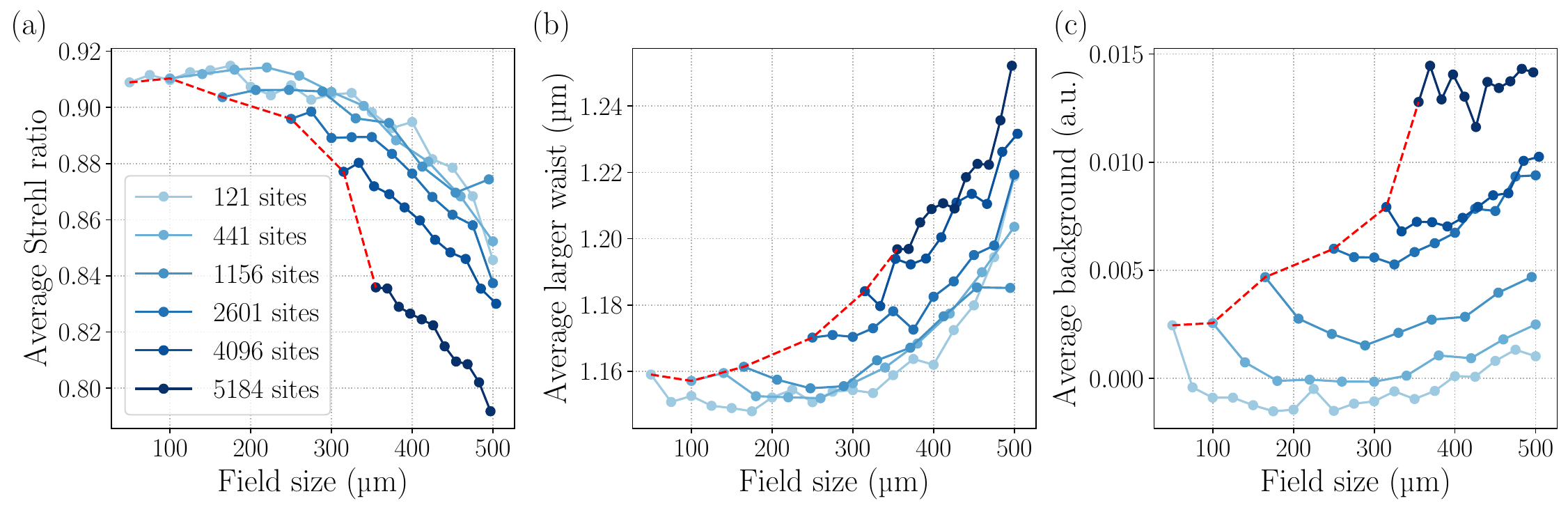}
\caption{Performance of the aberration correction as a function of the field size for a number of spots ranging from 121 to 5184. Lines connecting data points are a guide to the eyes. For each number of spots the smallest array corresponds to $d=5\,\mu$m. For the largest number of spots $d$ ranges from 5 to 7\,$\mu$m. (a), (b) and (c) present the variation of the average value of the Strehl ratio $\bar S$, maximum waist $\bar w_{\mathrm{max}}$ and background intensity $\bar I_{\mathrm{back}}$ respectively. The latter is normalized to the maximum intensity of the central spot highlighted by a circle on fig.\,\ref{Fig2}(g). A red dotted line connects the datapoints with $d=5\,\mu$m.}
\label{Fig5}
\end{figure*}

Finally, we study the quality of the corrected arrays as a function of their spatial extension and spot density for arrays containing from 121 to 5184 spots in a field of view ranging up to 500\,$\mu$m. For each number of spots the distance $d$ varies from 5 $\mu$m to the maximal value corresponding to an array size of 500\,$\mu$m.  As seen on fig\,\ref{Fig5}(a), the arrays with up to 1156 spots have an average Strehl ratio $\bar S$ above $0.9$ for fields up to $300\,\mu$m. The corresponding standard deviation of $\bar S$ is $<0.1$. We observe no more than 10\% decrease of $\bar S$ for field sizes larger than $300\,\mu$m.  For the explored range of arrays, the average larger waist [fig\,\ref{Fig5}(b)] only increases by about 10\% for the largest array with the highest spot density. We also observe that the homogeneity of power of individual spots degrades as we increase the size of the arrays and the number of spots. For the array with the highest number of spots (5184), the standard deviation of the power per spot varies from 5\% to 11\% when the array size increases from 350 to 500\,$\mu$m. The background intensity [fig\,\ref{Fig5}(c)] is not significant up to about 1000 sites in the arrays. It grows significantly for larger arrays when increasing the size of the field. For the largest array, the background intensity integrated over the $5\,\mu$m diameter region of interest used for the fit is on average 14$\%$ of the total power of the integrated Gaussian peak. The background intensity is related to a speckle background  clearly visible on fig.\,\ref{Fig2}(h). Its amplitude increases when spots are closer to the zeroth diffraction order of the SLM. It originates from the SLM imperfections, namely the limited number of pixels and the crosstalk between neighboring pixels, which become particularly important at the positions corresponding to $2\pi$ phase jumps between two consecutive pixels of the corrected phase mask. Note that aberration compensation also reduces the intensity of the background as seen by comparing fig.\,\ref{Fig3}(d) and (h).
This background significantly contributes to the distortion in shape and position of the individual spots. This is especially visible on the right side of arrays [Fig.\,\ref{Fig3}(d)] where we observe a larger ellipticity of spots with random orientation. 

\section{Array of bottle beams}

We now apply our field compensation method to arrays of Bottle Beams (BOBs). These arrays are particularly valuable for optically trapping Rydberg atoms, which are low-field seekers\cite{barredo_three-dimensional_2020,ravon_array_2023}. Similar BOBs, corrected using our method, have been used in\cite{ravon_array_2023} to trap Rydberg atoms over millisecond timescales. We prepare BOB arrays by adding a mask $\phi_\text{BOB}(\bm \rho)$ to the masks that generate the arrays of Gaussian spots. This added mask comprises two concentric areas with uniform phase and a $\pi$ relative phase-shift. For the same pupil and Gaussian beam waist on the SLM as above, the central region's radius is set to approximately 3.35 mm in order to transform every Gaussian spot into a BOB, precisely centered at the same position.

\begin{figure*}[tp]
\centering\includegraphics[width=13cm]{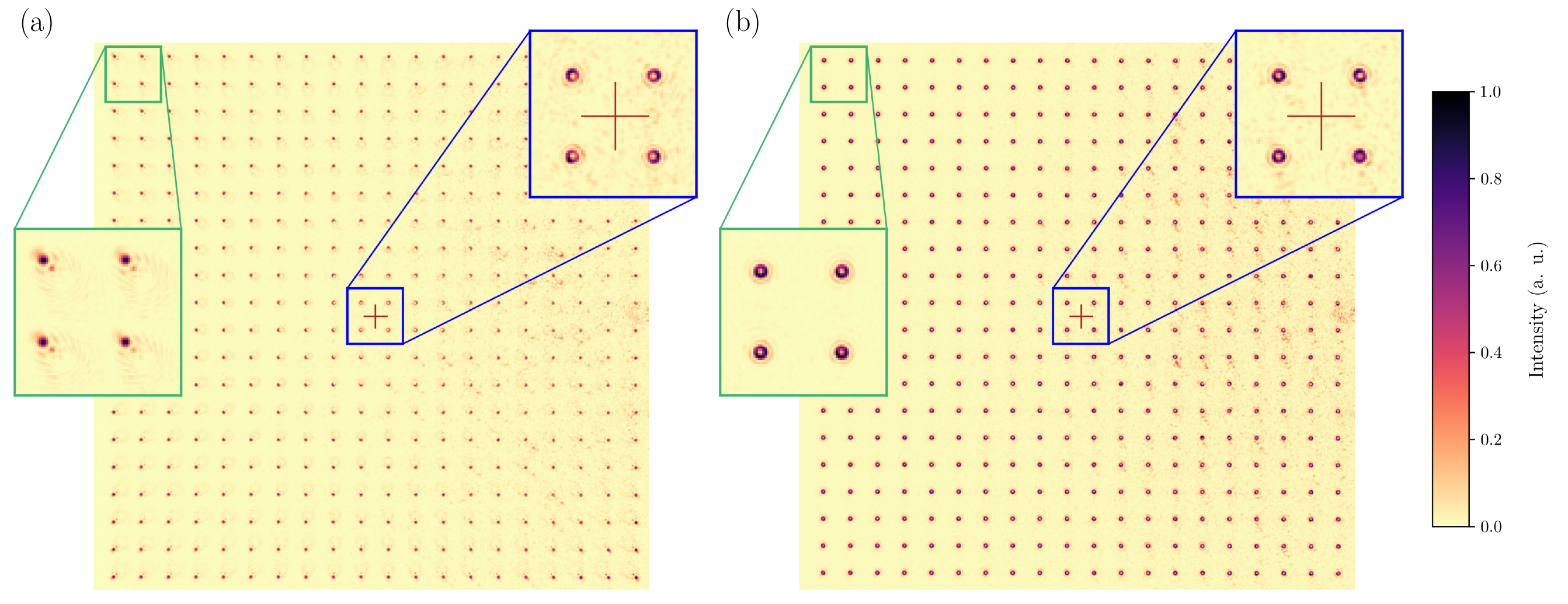}
\caption{Field correction of a $20\times 20$ array of BOBs with $d=25\,\mu$m. The array covers a field of $475\times475\,\mu$m. (a) and (b) are obtained without and with field correction respectively. In both cases, the mask $\phi_0(\bm \rho)$ is applied, correcting aberrations on the optical axis. Insets are zooms onto the central and top-left regions highlighted in blue and green, respectively. The red cross corresponds to the optical axis.}
\label{Fig6}
\end{figure*}

Without field correction [fig.\ref{Fig6}(a)], BOBs, which are much more sensitive to aberrations than Gaussian spots, open even at the minimal distance of $18\,\mu$m from the optical axis. At the array's periphery, the BOB's characteristic ring shape vanishes entirely, leaving only simple spots. In contrast, with field correction [fig.\ref{Fig6}(b)], all sites display the shape of a closed circle.

\section{Conclusion}

We have implemented a method to correct the aberrations introduced by a nearly centered optical system to produce aberration-free arrays of focal points. It uses holographic masks generated by a phase-only SLM. Adaptive optics methods compensate aberrations only close to a corrected spot, over a surface limited by the native FOV of the optical system. Our method, however, compensates the full accessible optical field. In addition, it is not limited to the generation of arrays of Gaussian tweezers. It produces high quality bottle beams, in spite of their critical sensitivity to aberrations. This is instrumental for the trapping of Rydberg atoms\cite{barredo_three-dimensional_2020, ravon_array_2023}. The method can be used for generating arrays of arbitrarily shaped spots. It can be extended directly to the generation of corrected three-dimensional arrays by adding a defocus term to the local correction\cite{lee_three-dimensional_2016, barredo_synthetic_2018}. High precision control methods of trap depth, ellipticity and position demonstrated in\cite{chew_ultraprecise_2024} can be implemented on the top of our correction method. In particular, the homogeneity of the power of individual spots could be improved using the intensity-equalization algorithm described in \cite{wang_feedback_2025}. In the context of neutral atom trapping, our method does not apply to correction of aberrations of the imaging system used to detect individual atoms. This effect could limit the maximal available atom number. However, a zonal adaptive optics approach\cite{park_large-field--view_2017-2} using another SLM can be implemented for improving the imaging system. Even if the images of the spots overlap, one can also use AI-enabled imaging techniques in order to improve the single-atom detection fidelity \cite{impertro_unsupervised_2023}. 
Our method would also directly improve the quality of images obtained by scanning confocal microscopy\cite{booth_adaptive_2002, helmchen_deep_2005, schermelleh_guide_2010}, or sub-wavelength Scanning Near-field Optical Microscopy (SNOM)\cite{hell_breaking_1994}. The method could be extended to the projection of arbitrary images instead of arrays of spots. This goes beyond the scope of this paper and will be the object of further studies.

As an illustration, we have applied our method to large-field aberration compensation of a low-cost aspherical lens, used out of its design wavelength range and presenting large imperfections. We have efficiently corrected the aberrations, extending its field of view from approximately 50$\,\mu$m to more than 500$\,\mu$m. The corrected field of view is limited by the maximum accessible diffraction angle with our SLM and can be straightforwardly expanded by using an SLM with a larger number of pixels. This performance, obtained using a single aspherical lens, is comparable to state-of-the-art performances of complex, bulky, high-numerical-aperture objectives\cite {manetsch_tweezer_2025}. The method is essentially limited by the SLM imperfections creating a background speckle, which interfere with individual spots and induces fluctuations of the shapes (amplitude and ellipticity) and positions of spots. Higher quality SLMs with less crosstalk between pixels would considerably reduce the background speckle. Alternatively, the speckle can be made negligible using high-resolution lithographically-fabricated holograms\cite{holmanTrappingSingleAtoms2024}. We anticipate that combining our CWGS hologram calculation method together with the use of lithographically fabricated holograms will simultaneously reduce spot fluctuations related to speckle and correct the residual aberration of the optical system.

Our work compensates field aberrations of a nearly centered optical system by introducing an efficient way to characterize aberrations in term of modified Seidel coefficients. We demonstrate that it is sufficient to introduce only four modified Seidel corrections to obtain diffraction-limited spots over the full accessible optical field. These corrections correspond to 8 parameters: coma of order 1 and 2, astigmatism and curvature and the centers of the latter two. They are determined by optical measurements performed at a small number of reference points exploring the field of view of interest. The knowledge of these coefficients enables the generation of arbitrary arrays with a large number of spots. Our method also applies to higher-order aberrations, corresponding to higher-order Zernike polynomials. Note that the method can also be used to generate 3D arrays of spots\cite{leonardo_computer_2007, lee_three-dimensional_2016, barredo_synthetic_2018}. 

Instead of using our model based on modified Seidel coefficient we could have optimized each spot of the array individually, based on a properly defined metric of the spot quality \cite{pozzi_anisoplanatic_2020}. Alternatively, aberrations can also be modeled using empirically defined aberration modes without a priori knowledge of the variation of aberrations within the field of view \cite{christen_full-volume_2025}. We think that by introducing a priori knowledge about aberration variation within the field of view of a nearly centered optical system, our model is more efficient in term of calibration overhead as compared to these more general approaches. Alternatively, one could also, in principle, use a real  image-based WGS algorithm in order to generate the target intensity profile. This method requires the calculation of images of large arrays with high resolution to resolve the shape of each individual spot. We anticipate that it will raise strong numerical efficiency issues as compared to our approach, which is based on compressed-sensing\cite{pozzi_fast_2018-1}. This approach only requires a number of evaluations of the field intensity equal to the number of spots instead of the calculation of the full intensity distribution at the focal plane of the lens using a 2D Fourrier-transform.

Finally, our method applies as well to any high-numerical-aperture optics with an optimized native large field in order to compensate for residual aberrations and further increase the aberration-compensated field of view. Our method thus drastically improves the useful field of any optical system used for generating large arrays of spots with foreseeable technological applications to high resolution imaging in biology by confocal microscopy, parallel laser machining of materials and 3D printing with optical resolution. 

\smallskip
\smallskip

%\begin{backmatter}
{\bf Acknowledgment} This publication has received funding by the France 2030 programs of the French National Research Agency (Grant No. ANR-22-PETQ-0004, project QuBitAF), under Horizon Europe programme HORIZON-CL4-2022-QUANTUM-02-SGA via the project 101113690 (PASQuanS2.1), by the European Union (ERC Advanced Grant No. 786919, project TRENSCRYBE). It has been supported by the Quantum Information Center Sorbonne as part of the program \emph{Investissements d'excellence} -- IDEX of the Alliance Sorbonne Université.

The authors declare no conflicts of interest.

Data underlying the results presented in this paper are not publicly available at this time but may be obtained from the authors upon reasonable request.

See Supplement for supporting content.
%\end{backmatter}

%%%%%%%%%% If using BibTeX:

\bibliography{bibliography}

\clearpage
\onecolumngrid

\section*{Supplemental document}
\setcounter{section}{0}

\section{Experimental details}

{\bf Optical setup}

We give here details about the optical setup. All optical elements before and after $\bf L$ were chosen carefully in order to minimize their aberrations. For imaging the SLM surface on the back focal plane of $\bf L$, we use a 4f afocal telescope made of  two lenses with 500\,mm focal length. Each lens is made of a pair of plano-convex lenses with a focal length of 1\,m, associated in doublet with the convex side nearly in contact as depicted on figure\,1 of the main text.

For reconstructing the phase mask $\phi_0(\bm \rho)$, the optical system $\bf I$  that images the focal plane of $\bf L$ on the camera combines an Olympus 100x, NA 0.9 microscope lens with a f=75 mm $\lambda/10$ plano-convex lens. It provides high resolution images (magnification $\times 41.7$) used for running the phase reconstruction algorithm presented in section 2 of this supplemental document. This imaging system is used close to its optical axis for imaging a single spot and introduces negligible aberrations.

For imaging arrays covering a large field, we use a $\times 50$ Leica lens (NA 0.5) followed by a $\times 10$ Mitutoyo lens (NA  0.26) providing a global magnification of $\times 5$. We carefully checked that it introduces negligible aberrations when imaging spots 250 $\mu$m away from the center of the array axis. To this end, we perform phase reconstruction for a given spot with the assembly of lenses placed at different transverse positions with respect to this spot. The reconstructed phase masks differ by less than 10\,mrad in the full field. From these measurements, we get the values of modified Seidel coefficients of this imaging system listed in table 1.
\begin{table}[h]
	\begin{center}
		\begin{tabular}{|c||c|c|c|}
			\hline
			\multirow{2}{*}{Aberration} & Imaging lenses & \multicolumn{2}{c|}{Aspherical lens} \\
			\cline{2-4}	 &\multicolumn{2}{c|}{Measured}& Theory\\
			\hline
			$1^{st}$ order coma --- $C^{'}$ (mrad$/\mu$m) & $\leq\SI{0.08}{} $ & $\SI{10.04}{}$ & $\SI{10.66}{}$\\
			$2{nd}$ order coma --- $C^{''}$ (mrad$/\mu$m) &  & $\SI{0.13}{}$ & $\SI{-0.20}{}$\\
			Astigmatism --- $A^{'}$ ($\mu$rad$/\mu$m$^{2}$) & $\leq\SI{0.3}{}$ & $\SI{3.3}{}$ & $\SI{3.5}{}$\\
			Field curvature --- $K^{'}$ ($\mu$rad$/\mu$m$^{2}$) & $\leq\SI{0.3}{}$ & $\SI{8.3}{}$ & $\SI{8.1}{}$\\
			\hline
		\end{tabular}
		\caption{Table of modified Seidel Coefficients values for the imaging system and for the aspherical lens. Theoretical values are obtained using OSLO software and using the theoretical shape of the aspherical lens provided by manufacturer.}
		\label{Table1}
	\end{center}
\end{table}

We use a Hamamatsu SLM (ref X13138) with $1272\times1024$ pixel. The pixel size is $12.5\,\mu$m. We use a pupil with a radius of 6\,mm. The maximal Nyquist-limited accessible diffraction angle corresponds to a spot located $535\,\mu$m away from the zero order reflexion in the focal plane of the aspherical lens.

\vspace{3mm}
{\bf Strehl ratio}

The Strehl ratio of a given spot is defined as the ratio of its peak intensity to the peak intensity of a diffraction-free reference spot with the same total energy. For the calculation of the Strehl ratio, due to the small number of camera pixels within a spot (see inset in figure\,3 of the main text) we do not determine its peak intensity using the brightest pixel. We rather fit each individual spot with a two-dimensional Gaussian function and with, as fitting parameters, the center position, the intensity at center, the larger waist, the ellipticity, the orientation of the major axis of the elliptical shape and a uniform background intensity. We also account for the finite size of the camera pixel in the fit procedure. To compute the Strehl ratio we take the maximal intensity of the fit. We normalize it to the total integrated power over a $8$ pixel diameter region of interest (ROI) centered on the considered spot of the image on the camera. 

To estimate the peak intensity of a diffraction-free reference spot, we first calculate its theoretical shape by Fourier Transform of the actual intensity distribution of the incoming beam on the SLM that we measured carefully. Note that the calculated shape of the reference spot is close to Gaussian. We then make the convolution of this theoretical shape with the camera pixel size and we normalize it with the total power measured with the camera for the considered spot over its ROI. We analyze this theoretical image with the same fit procedure as the actual spot images in order to determine its peak intensity. This analysis predicts a diffraction-limited waist of $1.09(1)\,\mu$m.

\section{Phase reconstruction}

The goal is to reconstruct the field amplitude $\sqrt{I(\bm \rho)} e^{-i\phi(\bm \rho)}$, which produces the intensity profile observed in the focal plane of the lens $\bf L$. In order to determine the wavefront  $\phi(\bm \rho)$, we use a phase reconstruction algorithm, which relies on the acquisition of several images in different planes located at positions $z_j$ before and after the focal plane of $\bf L$\cite{hanser_phase_2003}. The principle of this algorithm consists in using several images of the Fresnel diffraction pattern of the incoming amplitude. This set of images is used to uniquely reconstruct the non-diffracted complex field amplitude that generated it. This is a Gerchberg-Saxton type algorithm, which we shall refer to as the Gerchberg-Saxton Phase Retrieval (GS-PR) algorithm.

\begin{figure}[ht]
	\begin{center}
		\includegraphics[width=1\textwidth]{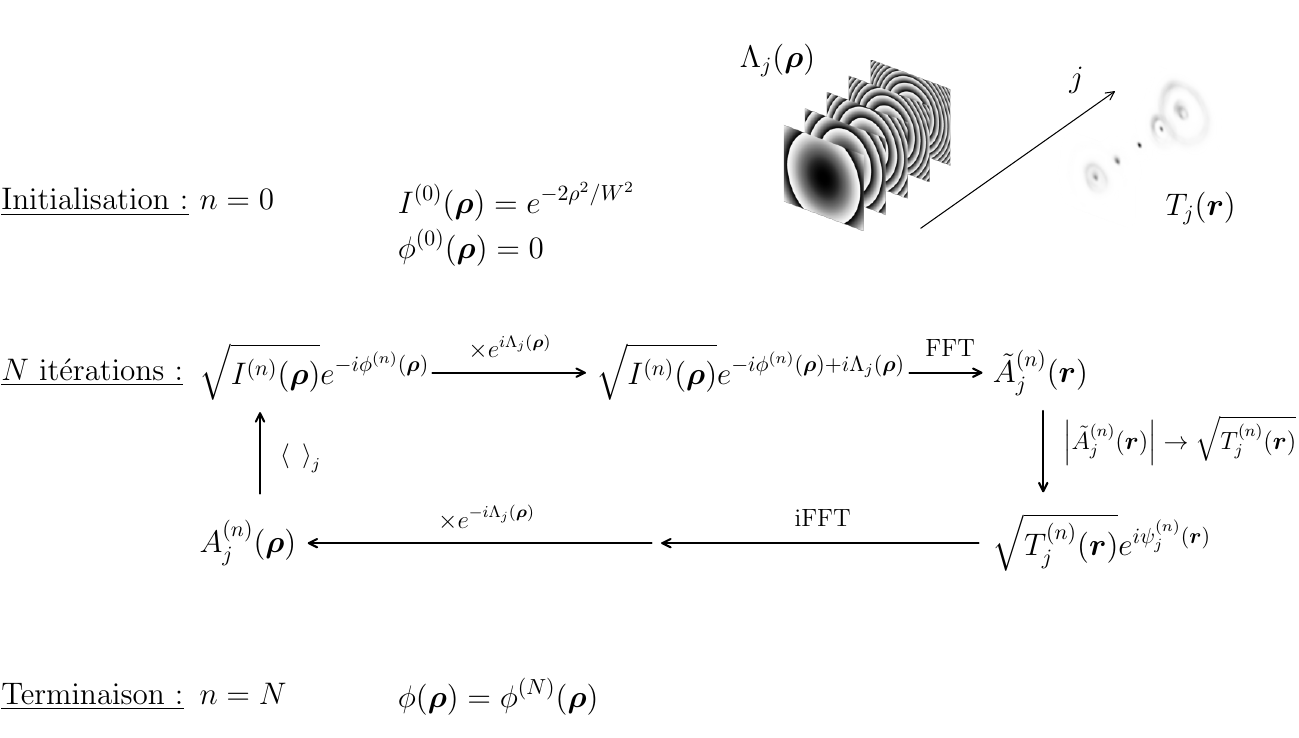}
	\end{center}
	\caption{Phase reconstruction GS-PR algorithm. }
	\label{fig_algo_aber}
\end{figure}

We run the GS-PR algorithm using images generated with the setup used for aberration correction and depicted in fig.\,\ref{Fig1}. In practice, instead of taking images with the camera at different positions $z_j$ with respect to the focal plane of $\bf L$, we equivalently take images after applying with the SLM defocus masks $\Lambda_j(\bm \rho)={k{\rho^2}\over{2f_j} }$. Each of these masks produces a shift $z_j= -f^2/(f+f_j)$ of the image along the optical axis. Before running the phase reconstruction algorithm, we take 15 images for different values of $z_j$ equally spaced in a $25\,\mu$m interval centered on $z=0$. It corresponds to a $\pm 4.5\, Z_R$ interval where $Z_R$ is the Rayleigh length of the focused beam. For each image, we get an intensity profile $T_j(\bm r)$. We then perform the iterative GS-PR algorithm as depicted on fig.\,S\ref{fig_algo_aber} with the following steps.

\begin{enumerate}[label=\arabic*.]

	\item {\bf Initialization}. We initialize the amplitude to be reconstructed in the back focal plane of $\bf L$ as $A^{(0)}(\rho)=\sqrt{I^{(0)}(\bm \rho)} e^{-i\phi^{(0)}(\bm \rho)}$, where $I^{(0)}(\bm \rho)=e^{-2\rho^2/W^2}$ is an initial Gaussian guess for the incident intensity. We start the algorithm with $\phi^{(0)}(\bm \rho)=0$.
	\item {\bf Iteration $n$}. 
	
	For each value of $z_j$ we perform the following operations:
	\begin{enumerate}[label=(\roman*)]
             \item Multiply the guess amplitude $A^{(n)}(\bm \rho)=\sqrt{I^{(n)}(\bm \rho)} e^{-i\phi^{(n)}(\bm \rho)}$ at step $n$ by $e^{i\Lambda_j(\bm\rho)}$ and compute its Fourier transform $\tilde A_j^{(n)}(\bm r)$. We define $\psi_j^{(n)}=\text{arg}[\tilde A_j^{(n)}(\bm r)]$.
             \item Replace the calculated amplitude $|\tilde A_j^{(n)}(\bm r)|$ by the measured amplitude $\sqrt{T_j(\bm r)}$ and compute the inverse Fourier transform of the  $\sqrt{T_j(\bm r)}{ e^{i \psi_j^{(n)}(\mathbf{r})}}$. We multiply it by $e^{-i\Lambda_j(\bm\rho)}$ and obtain a complex amplitude $A^{(n)}_j(\bm\rho)$.
            \item Compute the average over $j$ of all $A^{(n)}_j(\bm\rho)$. This average is the guess amplitude $A^{(n+1)}(\bm\rho)$ for the next iteration.
         \end{enumerate}
         \item {\bf Termination} after a fixed number of iterations $N$. 
\end{enumerate}

At the end of the algorithm the amplitude $A^{(N)}(\bm \rho)=\sqrt{I^{(N)}(\bm \rho)} e^{-i\phi^{(N)}(\bm \rho)}$ leads to an estimate of the intensity $I(\bm \rho)=I^{(N)}(\bm \rho)$ at the entrance pupil of $\bf L$ and of the phase distortion $\phi(\bm \rho)=\phi^{(N)}(\bm \rho)$ introduced by $\bf L$. We typically perform $N=30$ iterations for a single run of the GS-PR algorithm. To improve the convergence, we run the algorithm three times. Between two consecutive runs, we apply the correction mask determined by the previous run. The final correction mask is the sum of the three masks obtained in the three runs.
\begin{figure*}[t]
	\begin{center}
		\includegraphics[width=1\textwidth]{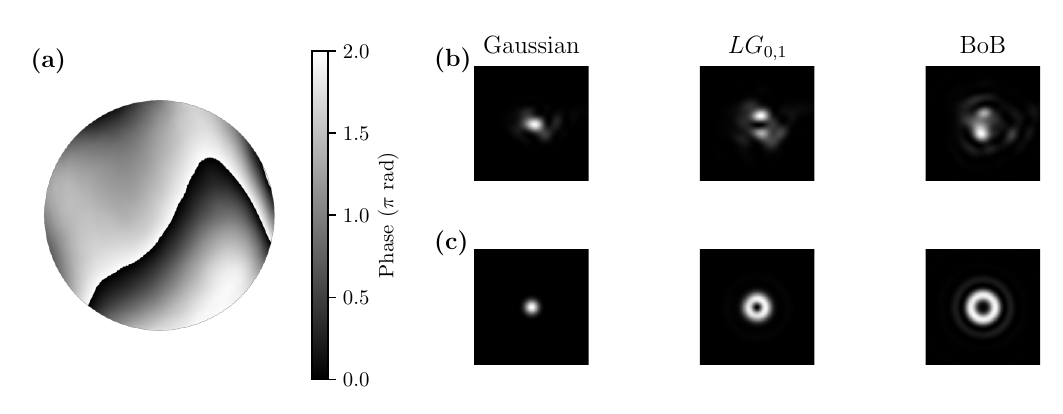}
	\end{center}
	\caption{{\bf Aberration correction}. (a) Measured phase distortion introduced by $\bf L$ on the optical axis. The circular region corresponds to the 12\,mm diameter pupil of the SLM. (b) Distorted images produced on the optical axis without aberration correction. The SLM phase mask is set to form a Gaussian, a Laguerre-Gauss ($LG_{0,1}$) or a bottle beam (left to right panels). (c) Same as (b) with aberration correction performed by subtracting with the SLM the phase distortion presented in (a).}
	\label{fig_GS_RP_images}
\end{figure*}

The wavefront determination accuracy was estimated by applying with the SLM a known phase distortion proportional to a given Zernike polynomial to a perfectly corrected spot. Comparison between applied and reconstructed distortion leads to an estimated 35\,mrad accuracy of the phase reconstruction method, corresponding to $ \lambda/180$. 

Figure\,\ref{fig_GS_RP_images}(a) shows the measured phase distortion introduced by our low-cost aspherical lens $\bf L$ on its optical axis. The standard deviation of the phase is 1.49 rad, corresponding to about $\lambda/4$. This large phase distortion results in a strong distortion of the observed images for various beam shapes as seen on fig.\,\ref{fig_GS_RP_images}(b). The corrected images shown on Figure\,\ref{fig_GS_RP_images}(c) are diffraction limited and do not present any significant distortion.

\section{Corrected array generation with the CWGS algorithm.}

We present here the principle of the CWGS algorithm that we use for the generation of corrected arrays.
In order to generate an array of spots with positions $\{\bm{r}_p\}_{p\in \{1,..., P\}}$, we search for a complex field amplitude with the form 
\begin{equation}
a(\boldsymbol{\rho}) = \frac{\sum_p w_p e^{i \frac{k}{f} \mathbf{r}_p \cdot \boldsymbol{\rho}} \,\, e^{-i \Phi({\bm r}_p, {\bm \rho})}}{\left| \sum_p w_p e^{i \frac{k}{f} \mathbf{r}_p \cdot \boldsymbol{\rho}} \,\, e^{-i \Phi({\bm r}_p, {\bm \rho})}\right|}
\end{equation}
at the back focal plane of $\bf L$. The argument of $a(\boldsymbol{\rho})$ will define the phase pattern to apply with the SLM to the incoming plane wave in order to generate the desired array of corrected spots in the focal plane of  $\bf L$. Note that this expression of $a(\boldsymbol{\rho})$ has by definition a constant unit modulus and corresponds to phase-only wavefront modulation. The algorithm determines the complex weights $\{w_p\}_{p\in \{1,..., P\}}$ that generate the desired array of spots by iteratively applying the following operations : 

\begin{figure}[h]
	\begin{center}
		\includegraphics[width=1\textwidth]{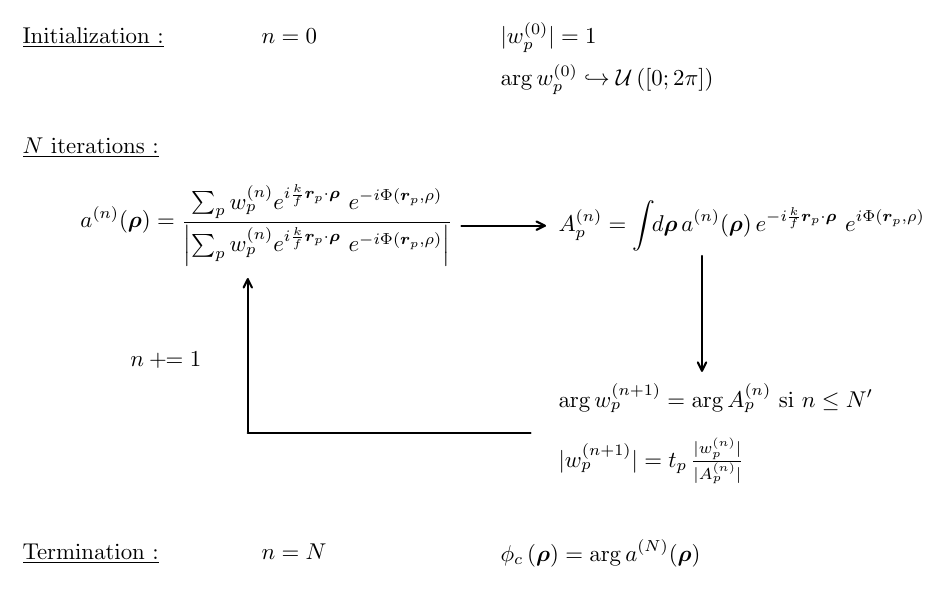}
	\end{center}
	\caption{Principle of the CWGS algorithm.}
	\label{fig_algo_field}
\end{figure}

\begin{enumerate}[label=\arabic*.]

	\item {\bf Initialization}. We initialize the amplitude to  $a^{(0)}(\boldsymbol{\rho})$ using weights $w_p^{(0)}$ with unit moduli and random phases.   
	
	\item {\bf Iteration $n$}. 
	
	\begin{enumerate}[label=(\roman*)]
             \item Using the amplitude $a^{(n)}(\boldsymbol{\rho})$, we compute 
		\begin{equation}
    			A_{p}^{(n)} = \int d\boldsymbol{\rho} \, a^{(n)}(\boldsymbol{\rho}) \, e^{-i \frac{k}{f} \mathbf{r}_p \cdot \boldsymbol{\rho}} \,\, e^{i \Phi({\bm r}_p, {\bm \rho})},
			\label{Ap}
		\end{equation}
		for each value of $p$. It corresponds to the complex amplitude of the field at position ${\bm r}_p$ in the focal plane of $\bf L$, including the aberration introduced by this lens. Note that due to the ${\bm r}_p$ dependency of $\Phi({\bm r}_p, {\bm \rho})$, this amplitude is not the Fourier transform of $a^{(n)}(\boldsymbol{\rho})$.
             \item From eq.\,\ref{Ap}, one gets the updated value of the weight $w_p^{(n+1)}$ for each value of $p$ according to the equations
             	\begin{align}
   		 \arg w_{p}^{(n+1)} &= \arg A_{p}^{(n)}, \\
   		 |w_{p}^{(n+1)}| &= t_{p} \frac{|w_{p}^{(n)}|}{|A_{p}^{(n)}|}.
		\end{align}
	The choice of the coefficients $t_p$ allows one to eventually tune the relative intensities of individual spots.
         \end{enumerate}
         \item {\bf Termination}. We stop after a fixed number of iterations $N$. Note that in order to ensure correct convergence of relative spot intensities, we no longer update the phase of $w_p^{(n)}$ after $N'$ iterations, with $N'<N$. The phase mask leading to the formation of corrected spots is defined as the argument of $a^{(n+1)}(\boldsymbol{\rho})$.
\end{enumerate}
   
The algorithm typically converges with $N=40$ and $N'= 10$. Note that if we neglect the field-dependent correction term $\Phi({\bm r}_p, {\bm \rho})$ in the CWGS algorithm, we recover the standard WGS algorithm widely used to generate arrays of discrete spots. We implemented the algorithm using a Nvidia Quadro P2000 GPU. The computational time is 2\,s for generating the 34*34 corrected array shown in figure\,3 in the main text. There is no significant overhead as compared to running the standard WGS algorithm for generating a similar, uncorrected array.

\section{Determination of modified Seidel coefficients}

We give here the complete expression of the functions used to fit the measured $\theta_n^m({\bm r}_q)$ in order to determine the modified Seidel coefficients

\begin{equation}
	\begin{split}
	\theta_3^1(x,y) &= C^{\prime} x\\
	\theta_3^{-1}(x,y) &= C^{\prime} y\\
	\theta_5^1(x,y) &= C'' x\\
	\theta_5^{-1}(x,y) &= C'' y\\
	\theta_2^2(x,y) &= A^{\prime} [(x-x_0^A)^2-(y-y_0^A)^2]\\
	\theta_2^{-2}(x,y) &= A^{\prime} 2(x-x_0^A)(y-y_0^A)\\
	\theta_2^0(x,y) &= K^{\prime} [(x-x_0^K)^2+(y-y_0^K)^2].
	\end{split}
\end{equation}
For each value of $n$ corresponding to two $\theta_n^m(x,y)$ functions with opposite values of $m$, we performed a simultaneous fit of the two corresponding datasets with the modified Seidel coefficients and centers as shared parameters. For the coma coefficients $C'$ and $C''$, we do not introduce center in the fit as the coma is already compensated on the optical axis by the application of the global compensation mask $\phi_0(\rho)$.

\end{document}